\documentclass[12pt]{ar}
\usepackage[english]{babel}
\usepackage{graphicx}

\onecolumn
\begin{document}

\title{Detection of the spectral binary (SB2) nature of BD${-6\degr}$1178}

\author{V.G.\,Klochkova \& E.L.\,Chentsov}

\date{\today}	     

\institute{Special Astrophysical Observatory RAS, Nizhnij Arkhyz,
             369167 Russia} 

\abstract{\mbox{BD$-6\degr$1178} identified with the infrared source
IRAS\,05238$-$0626 is shown for the first time to be a spectroscopic
binary (SB2) by analyzing the high-resolution spectra taken with the NES
echelle spectrograph of the 6-m telescope. The components of the binary
have close spectral types and luminosity classes: F5\,IV--III and F3\,V.
The heliocentric radial velocities are measured for both components at four
observing moments in 2004--2005. Both stars have close rotation
velocities, which are equal to 24 and 19\,km/s. We do not confirm the
classification of BD$-6\degr$1178 as a supergiant in the transition stage
of becoming a planetary nebula. BD$-6\degr$1178 probably is a young pre-MS
stars. It is possibly a member of the 1c subgroup of the Ori\,OB1
association.}

\titlerunning{\it Spectroscopic binary BD${ -6\degr}$1178}
\authorrunning{\it Klochkova \& Chentsov}

\maketitle

\section{Introduction}

In this paper we continue to publish the results of our spectroscopy of
stars with IR excesses (see [\cite{Klochkova1995, KSPV, IRAS23304,
AFGL2688, IRAS20000}] and references therein for the main results).
BD$-6\degr$1178 is an optical component of the infrared source
IRAS\,05238$-$0626 (with galactic coordinates l=208.9$\degr$,
b=$-21.8\degr$). This object is considered to be a candidate to
protoplanetary nebula (PPN) according to the observed excess of radiation
in the 12--60\,$\mu$m wavelength region and its position on the IR
colour--colour diagram [\cite{Garcia-1990, Reddy, Fujii}]. Recall that
according to modern concepts (see, e.g., [\cite{Block}]), objects observed
at the short-lived evolutionary stage of a young planetary
(protoplanetary) nebula are intermediate-mass stars evolving away from the
asymptotic giant branch (AGB) toward the stage of a planetary nebula. The
initial main-sequence (MS) masses of these stars lie in the
3--8\,${\mathcal M}_{\odot}$ mass interval. During the AGB stage these
stars have lost much mass in the form of a powerful stellar wind, and as a
result, at the PPN stage the stars have the form of degenerate
carbon-oxygen nuclei with typical masses of about 0.6\,${\mathcal
M}_{\odot}$ surrounded by expanding gas-and-dust shell. The astronomers
are interested in studying PPNs, first because they allow one to study
stellar-wind driven mass loss and second, because they offer a unique
opportunity of observing the result of the stellar nucleosynthesis,
mixing, and dredge-up of products of nuclear reactions that occurred
during the preceding evolution of the star.

About a dozen objects overabundant in heavy metals synthesized via
neutronization of iron nuclei under the conditions of low neutron density
(\mbox{$s$-process}) have been found among the PPN-candidate studied. An
analysis of the properties of PPNs showed that the expected overabundances
of $s$-process elements are observed only the atmospheres of \mbox{C-rich}
stars whose IR spectra contain an emission at 21\,$\mu$m
[\cite{Klochkova1995, IRAS20000, Klochkova1997, Winckel}]. However, the
overwhelming majority of PPNs exhibit neither carbon (\mbox{O-rich} stars)
nor heavy-element overabundance (see, e.g., [\cite{Klochkova1995, QYSge,
IRAS19475}]). The correlation found between the excess of heavy elements
in the star's atmosphere and the peculiarity of the IR spectrum of the
envelope of the star remains unexplained and hence a further increase of
the sample of PPN objects studied is needed.

Currently, we know little about BD$-6\degr$1178. Its sky coordinates for
the epoch of 2000 are: $\alpha$=05$^{\rm h} 26^{\rm m} 19.8^{\rm s}$,
$\delta$=$-6\degr 23^{\rm '} 57^{\rm ''}$. The $V$- and $B$-band apparent
magnitudes are equal to $V$\,=\,$10.52^m$ and
\mbox{$B$\,=\,$10.96^m$ [\cite{Fujii}]}. Some evidence for the photometric
variability of the star was found: according to the NSVS catalog
[\cite{Wozniak}], the mean magnitude of the star in a close-to-$R$-filter
passband varies in the interval 10.78--$10.87^m$ and its standard
deviation is about $0.01^m$. Modeling of the spectral energy distribution
based on the multicolor photometry in the visual and near IR yields
effective-temperature values T$_{eff}$ ranging from 8000\,K
[\cite{Garcia-1990}] to 7400\,K [\cite{Fujii}], which corresponds to late
A --- early F-subclasses.

As for spectroscopic observations, only low-resolution
($\approx$5\,\AA/pixel) spectra have been published so far for
BD$-6\degr$1178. These spectra yielded the following estimates for the
spectral type: F2\,II [\cite{Reddy}], F4 [\cite{Suarez}], and F5
[\cite{Torres}]. In view of the aforesaid, it becomes evident from these
results that further detailed study of the optical spectrum of the star is
needed. In this paper we report the results of our numerous
high-resolution spectroscopic observations of BD$-6\degr$1178 made with
the 6-m telescope of the Special Astrophysical Observatory of the Russian
Academy of Sciences (SAO RAS). The aim of this study is to perform
two-dimensional spectral classification, search for spectroscopic
variability, analyze the velocity field in the star's atmosphere and
envelope, and refine its evolutionary status. In Section\,\ref{observ} the
methods of observation and reduction are described; in
Section\,\ref{results} we present and analyze the observational data
obtained, and in Section\,\ref{conclus} we briefly sums up the main
results.

\section{Observations and analysis of the spectra}\label{observ}

We obtained our spectral material for BD$-6\degr$1178 using NES echelle
spectrograph [\cite{nes}] mounted in the Nasmyth focus of the 6-m
telescope of the SAO RAS. Observations were performed with a large-format
2048\,$\times$\,2048 CCD and with an image slicer [\cite{nes}]. The
spectroscopic resolution is equal to 60000. We use the modified
[\cite{Yushkin}] ECHELLE context of MIDAS package to extract the data from
two-dimensional echelle spectra. We remove cosmic-ray hits using median
averaging over two spectra taken in succession. Wavelength calibration is
performed using the spectra of a hollow-cathode Th-Ar lamp. We computed
the radial and rotational velocities obtained from these spectra by
superimposing direct and mirror-reflected line profiles and listed them in
Table\,\ref{data}. We controlled the instrumental mismatch between the
spectra of the star and those of the hollow-cathode lamp by the O$_2$ and
H$_2$O telluric lines. Residual systematic errors do not exceed the
measurement errors (about 1\,km/s for a single line).

\section{Results and discussion}\label{results}

\subsection{Spectroscopic Binarity}

\begin{figure}[hbtp]
\includegraphics[angle=0,width=0.9\textwidth,bb=40 70 550 780,clip]{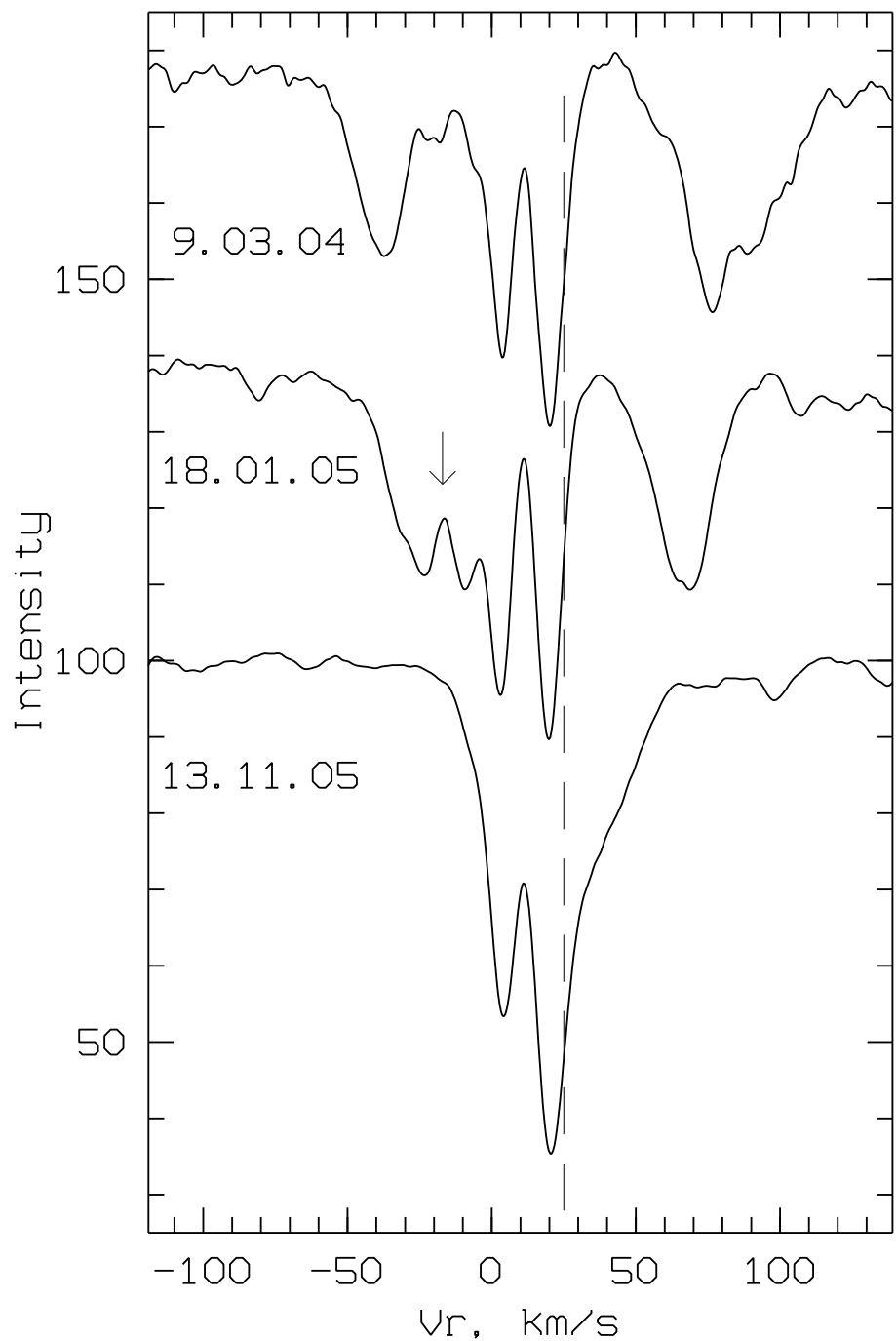}
\caption{Comparison of the spectra of  BD$-6\degr$1178 in the vicinity of
         the D2\,NaI line. The arrow on the spectrum of  18.01.05 indicates
	 a telluric emission feature. The vertical dashed line indicates
	 the adopted systemic velocity, \mbox{V$_{sys}\approx$25\,km/s.}}
\label{NaD}
\end{figure}

Our main result is that we are the first to find that BD$-6\degr$1178 is a
spectroscopic binary. This is a double-line spectroscopic binary with
narrow and easy-to-separate lines. It is evident from Table\,\ref{data}
that the maximum offset between the spectra of the components we recorded 
is equal to about 120\,km/s, which exceeds the line width at least by a
factor of five. The two companions have rather close depths of absorptions
in their spectra and hence rather close spectral and luminosity classes.
We consider the primary companion to be the star with slightly deeper and
wider absorption features (their central depths, widths, and equivalent
widths exceed the corresponding parameters of the absorption features of
the secondary companion on the average by 8\%, 25\%, and 35\%,
respectively). Figure\,\ref{NaD} shows the fragments of the spectra in the
vicinity of the D2\,NaI line for various observational dates. The
heliocentric radial velocities of the interstellar components are equal to
Vr\,=\,4 and 20\,km/s.

\begin{figure}[hbtp]
\includegraphics[angle=-90,width=1.0\textwidth,bb=50 110 520 780,clip]{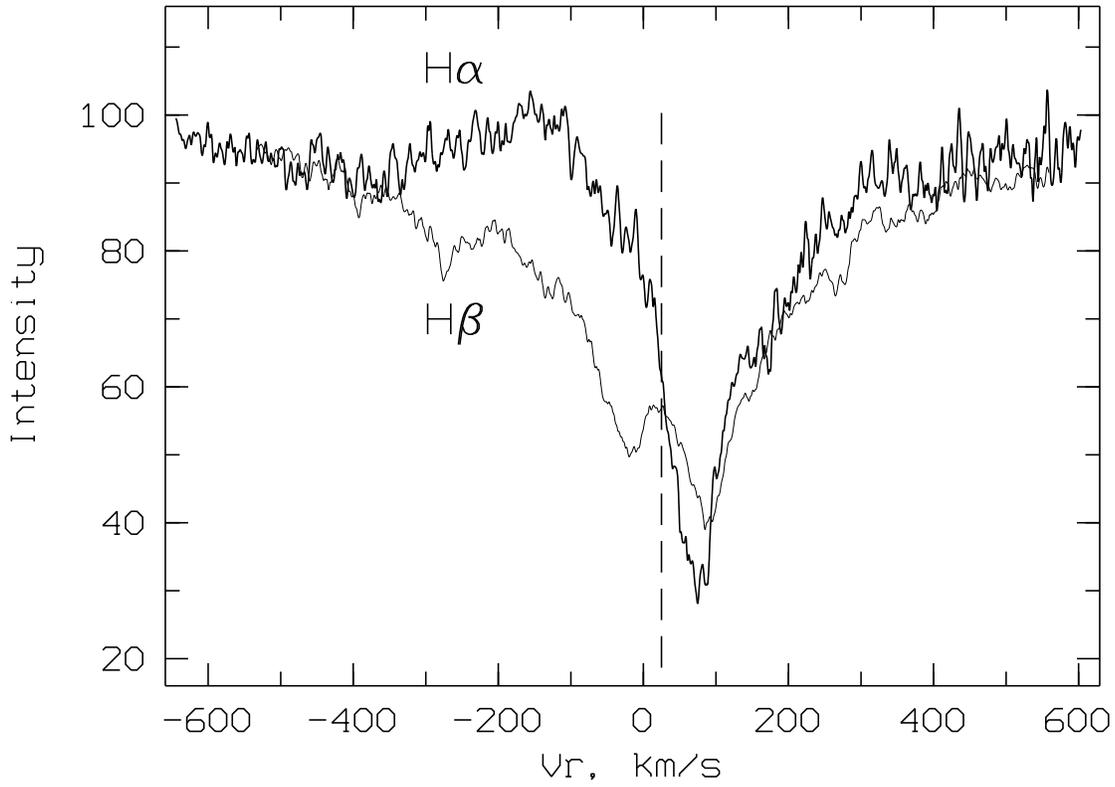}
\caption{Profiles of the H$\alpha$ (9.03.04) and H$\beta$ (24.09.05,
        the thin line) lines. The vertical dashed line shows the adopted
	systemic velocity of  V$_{sys}\approx$25\,km/s.}
\label{profiles}
\end{figure}

Differential shifts can be seen only for the H$\beta$ and H$\alpha$ lines.
For H$\beta$ the shifts are small and they may be due to the blending of
wide components; in the spectrum taken on 24.09.05 the velocities of the
H$\beta$ absorption components are equal to 87 and $-$20\,km/s, and the
spectrum taken on 13.11.05 the velocity of the core of the unsplitted
absorption is equal to 26\,km/s. In case of H$\alpha$ splitting disappears
--- possibly because of the more complex shape of the component profiles.
In the spectral interval recorded in our observations it is the only line
where emission is possible. Otherwise it is difficult to explain the
well-marked difference between the profiles of H$\alpha$ and H$\beta$ (see
Fig.\,\ref{profiles}), which differ little from each other in the spectra
of normal F-type stars (as it is evident, e.g., from high-resolution
spectral atlases [\cite{Montes, UVatlas}].

We have no spectra containing both these lines, but we can compare their
profiles in close phases: the H$\alpha$ profile of 9.03.04 and the
H$\beta$ profile of 24.09.05. In Fig.\,\ref{profiles} the two profiles
are superimposed and the vertical line indicates the adopted
$\gamma$-velocity of the system (${\rm V_{sys}\approx}$25\,km/s). Both
components show up conspicuously on the H$\beta$ profile of 24.09.05 and
their contribution to the total profile differ more than the corresponding
contributions for other lines. The first profile (Vr\,=\,87\,km/s) is
deeper than the second one (Vr\,=\,--20\,km/s) by 22\%. The H$\alpha$
profile of 9.03.04 is asymmetric: absorption in its red half is much
stronger than in the blue half. Only the first component (its radial
velocity is 75\,km/s) can be clearly seen, it is deeper than the
corresponding component of H$\beta$ line; the second component possibly
contains emission, which rises to the continuum level at
Vr$\approx$--130\,km/s.

\begin{figure}[hbtp]
\includegraphics[angle=-90,width=0.9\textwidth,bb=35 60 560 780,clip]{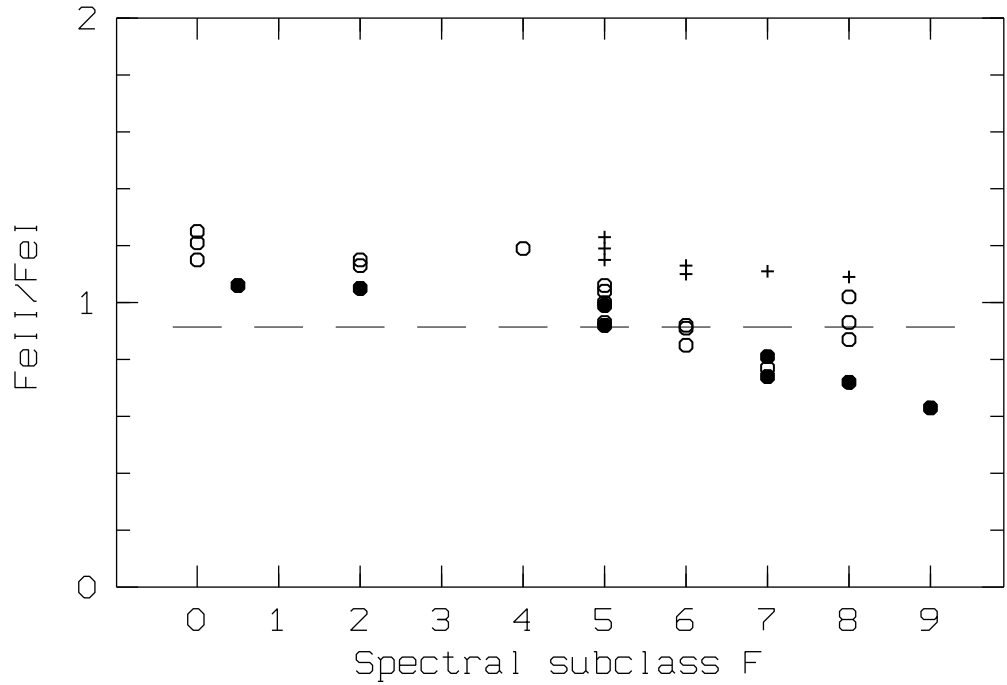}
\caption{Dependence of the ratio of central absorption depths,
         R(FeII\,4924)$/$R(FeI\,4921), on spectral type and
         luminosity class. The filled circles, open circles,
	 and crosses correspond to luminosity classes V, IV-III,
	 and II-I, respectively. The dashed line shows the depth ratio
	 for BD$-6\degr$1178 at the phase of component conjunction
         on 13.11.05.}
\label{Calibr}
\end{figure}

The two-dimensional spectral classification of F-type stars is rather
complex. Our observational material can be used to perform it by comparing
the intensities of the lines of neutral metals and their ions. We
constructed the calibrating curves for the line pairs
(FeII\,4731/FeI\,4737, FeII\,4924/FeI\,4921, et al.) using the
high-resolution spectra from the atlas of Klochkova et
al.~[\cite{UVatlas}] and of the ELODIE.3 library [\cite{Prugniel}].
Figure\,\ref{Calibr} shows, by way of an example, the dependence of the
ratio of the central depths of absorptions on spectral type and luminosity
class for the FeII\,4924\,\AA{} and FeI\,4921\,\AA{} lines pair.

The FeII/FeI absorption depth ratios, like those found in the spectrum of
BD$-6\degr$1178, are observed in the interval of MK--classes from F4V to
G0I. However, the color index of BD$-6\degr$1178
(B$-$V$\approx$0.44--0.47) restricts the spectral type of the star: it
cannot be later than F6\,V--III or F7\,II--I, and therefore the supergiant
option has to be rejected. This is also evident from the low-resolution
spectrum of BD$-6\degr$1178 reported by
Reddy~and~Parthasa\-rathy~[\cite{Reddy}]: the blend of the IR oxygen
triplet OI\,7774\,\AA{} in this spectrum is much weaker than in the
spectra of F-type supergiants. Our spectrum taken on 13.11.05 (at a phase
close to the conjunction of the components) yields a mean spectral type of
F5\,IV for the entire system, other spectra yield somewhat later type and
higher luminosity for the primary compared to the secondary: F5\,IV--III
and F3V, respectively. Interstellar extinction is weak, and that's the
case not only for BD$-6\degr$1178 (it does not exceed 0.15$^{m}$), but
also for the neighboring stars. With interstellar extinction and the
binary nature of the star taken into account, its heliocentric distance
can be estimated at about 450\,pc.

\begin{figure}[hbtp]
\hspace{-1.0cm}
\centerline{\vbox{
\includegraphics[angle=-90,width=0.8\textwidth,bb=50 100 560 780,clip]{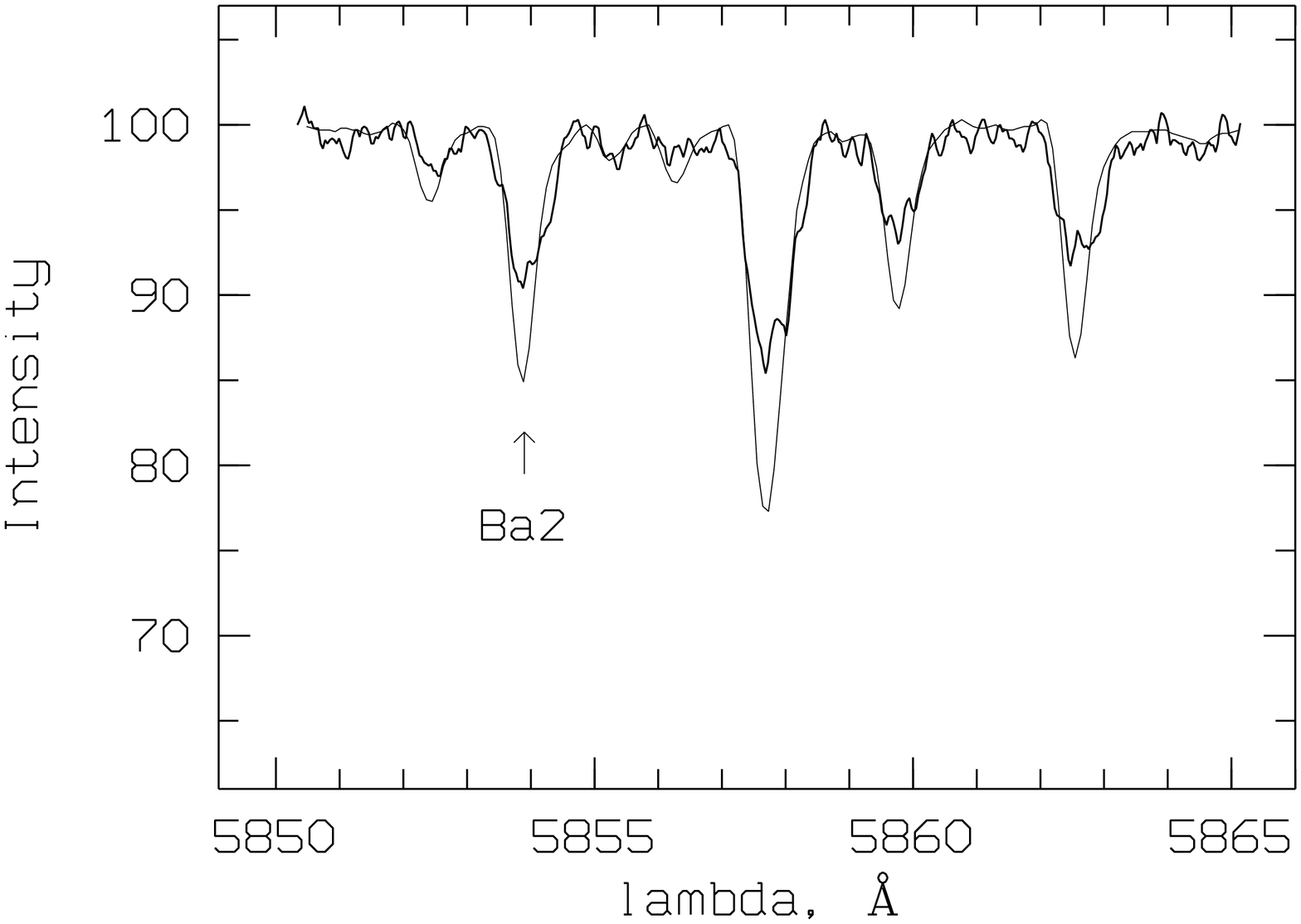}
\includegraphics[angle=-90,width=0.8\textwidth,bb=50 100 560 780,clip]{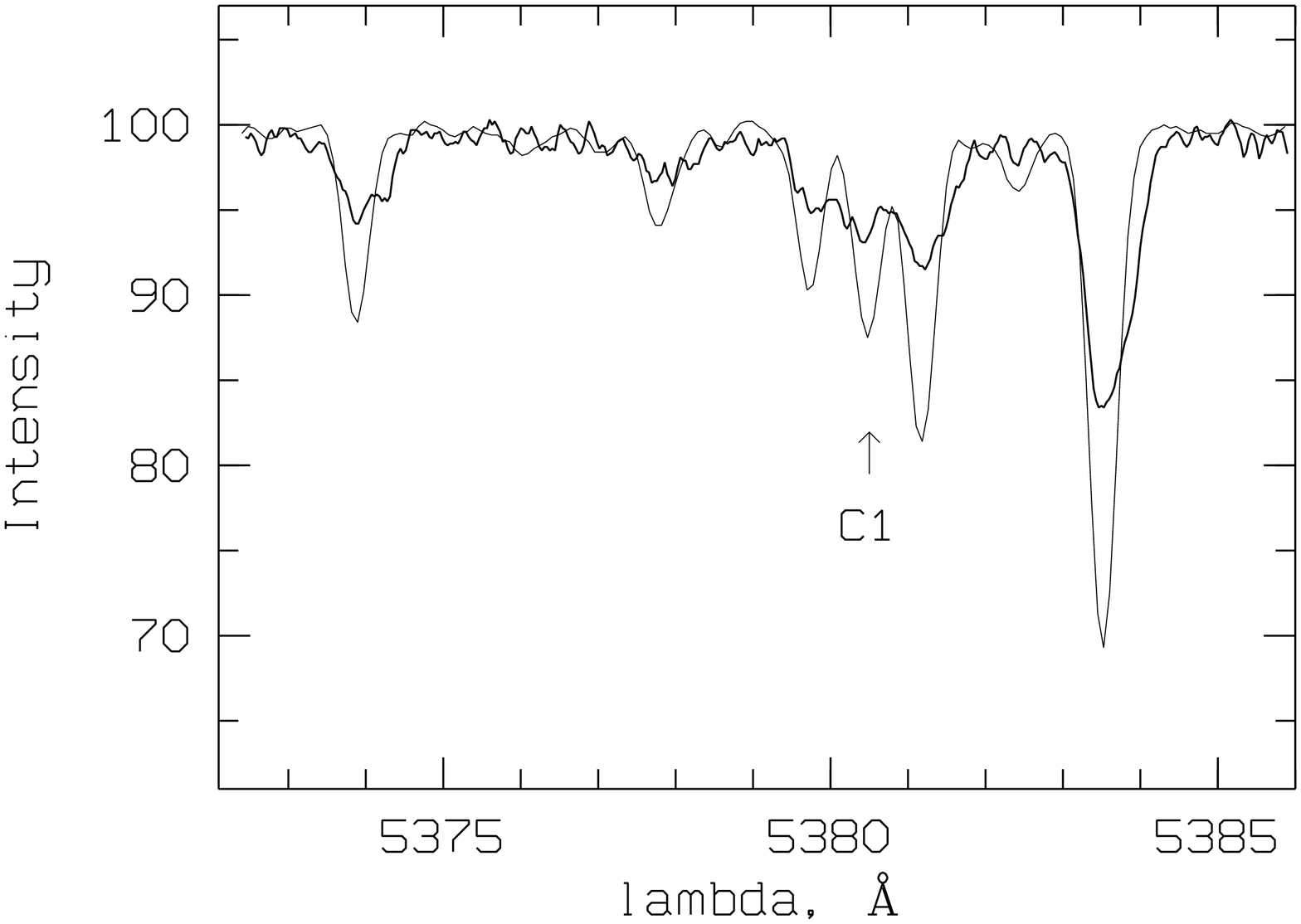}}}
\caption{The BaII\,$\lambda$5853\,\AA{} and CI\,$\lambda$5380\,\AA{}
         lines in the spectra of BD$-6\degr$1178 (13.11.05, thick line)
	 and Procyon (the thin line).}
\label{Barium}
\end{figure}

One would expect the spectrum of a post-AGB supergiant to exhibit
anomalous equivalent widths for the chemical elements whose abundances are
subject to change in the course of stellar evolution. This concerns, first
and foremost, elements of the CNO group and heavy metals whose nuclei are
synthesized during slow neutronization (Sr, Y, Zr, Ba). However, we found
no important differences between the spectrum of BD$-6\degr$1178 and those
of unevolved stars of similar spectral types. To illustrate this point, in
Fig.\,\ref{Barium} we compare fragments of the spectra of BD$-6\degr$1178
and Procyon (F5IV--V) containing the BaII\,$\lambda$5853\,\AA{} and
CI\,$\lambda$5380\,\AA{} lines. We thus obtain an additional corroboration
for our spectral classification of \mbox{BD$-6\degr$1178} as a system
consisting of low-luminosity stars.

\subsection{The Evolutionary Status of {\rm BD}$-6\degr${\rm 1178}}

We already mentioned the scantiness of published data for BD$-6\degr$1178.
We now also point out the inconsistency of published estimates of the
distance to the star and its evolutionary status: Fujii et al.
[\cite{Fujii}] classify BD$-6\degr$1178 as a post-AGB star at a distance
of about 10\,kpc, whereas Suarez et al. [\cite{Suarez}] classified it as a
young star based on optical low-resolution spectra. Our estimate of the
distance to this pair---about 450\,pc---agrees with the results of
Suarez~et~al.~[\cite{Suarez}].

Garcia-Lario et al. [\cite{Garcia-1997}] obtained and analyzed near-IR
photometry for an extensive sample of 225 sources including
IRAS\,05238$-$0626. They found that in the (H-K, J-H) color--color
diagram IRAS\,05238$-$0626 lies in a domain populated mostly by
post-AGB stars. However, this domain may also contain young stars of the
T\,Tau and Herbig's Ae/Be types. This result led the above authors to
conclude that the evolutionary status of IRAS\,05238$-$0626 is uncertain:
it may be either post-AGB or T\,Tau star.

In Table\,4 of Fujii et al. [\cite{Fujii}], which gives the parameters of
26 candidate to PPN objects, IRAS\,05238\,$-\,$0626 source stands out
because of its rather high effective temperature and low mass-loss rate.
The latter fact is inconsistent with the source's location in domain IV on
the van der Veen-Habing IR color-color diagram~[\cite{Veen}]. The above
authors~[\cite{Veen}] define this domain as the locus of variable stars
with high mass-loss rate and producing a powerful circumstellar shell.
Moreover, IRAS\,05238\,$-\,$0626 also differs from typical PPN by its low
IR flux. The $\lambda$\,12$\mu$m flux of this source is F$_{12}$=0.59\,Jy,
which is  one to one-and-a-half orders of magnitude 
lower than the corresponding fluxes of such well  studied
post-AGB objects as IRAS\,04296+3429  (F$_{12}$=12.74\,Jy),
IRAS\,23304+6147 (F$_{12}$=8.56\,Jy), and IRAS\,07331+0021
(F$_{12}$=15.32\,Jy).

Reddy and Parthasarathy [\cite{Reddy}] analyzed a sample of 14 candidate
PPN objects including BD$-6\degr$1178 and concluded that BD$-6\degr$1178
is a highly evolved post-AGB star where a cooled-down remnant of the
circumstellar shell can be observed. The above authors adopted a
stellar-core mass typical of post-AGB stars---0.6\,${\mathcal
M}_{\odot}$---to obtain a high luminosity estimate $\lg(L/L_{\odot})=3.79$
and, consequently, large heliocentric distance (7\,kpc) for the star. Note
that Reddy and Parthasarathy~[\cite{Reddy}] pointed out the absence of
sources of molecular emission to be associated with BD$-6\degr$1178,
whereas post-AGB stars are typically characterized by thermal and maser
emission of CO, SiO, H$_2$O, and other molecules (see references in the
reviews by Kwok [\cite{Kwok}] and Klochkova [\cite{Klochkova1997}]). For
example, the IR sources IRAS\,04296+3429 and 23304+6147 mentioned above
are powerful sources of CO emission [\cite{Hrivnak}]. Moreover, the
presence of molecular features, which replace each other in the course of
evolution of the PPN, allowed Lewis [\cite{Lewis}] to trace the
chronological sequence of molecular spectra in stars at different stages
of post-AGB evolution.

In view of the facts mentioned above and the results of our spectral
classification, we suggest that the spectral binary BD$-6\degr$1178
(F5\,IV--III\,+\,F3V) may be a young object of the Galactic disk. Note
that its coordinates and its heliocentric distance of about 450\,pc allow
us to suspect that BD$-6\degr$1178 may be a member of the Ori\,OB1
association. According to the list of distances to stellar associations
based on Hipparcos data [\cite{Zeeuw}], subgroup 1c in the Ori\,OB1
association is located at a heliocentric distance of 506$\pm$37\,pc. Note
also that the adopted ${\rm V_{sys}}$ and measured velocities for the NaI(1)
interstellar-line components are consistent with the velocities
(15--28\,km/s) adopted from the SIMBAD database for 19 stars located
within $2.5\degr$ of BD$-6\degr$1178 within the heliocentric-distance
interval 0.2--0.7\,kpc.

Earlier Torres et al. [\cite{Torres}] while analyzing a sample of
candidate to T\,Tau-type stars included BD$-6\degr$1178 to their sample.
They analyzed the spectroscopic and photometric data and found a total of
17 and 13 T\,Tau and Herbig's He/Be stars, respectively. However, the
above authors could not place BD$-6\degr$1178 into either of the two
groups and classified it as ``miscellaneous''-type object. On the whole,
we see no reasons to classify \mbox{BD$-6\degr$1178} as a post-AGB star.
Note also that the detection of the spectroscopic binarity of SB2-type
provides further evidence to doubt the classification of BD$-6\degr$1178 as
a post-AGB star because there are no SB2-type binaries among the known
stars at the evolutionary stage considered. A number of post-AGB stars are
SB1-type binaries. The nature of the unseen companion is unknown, because
its spectral features do not show up in the spectra of post-AGB binaries.
HD\,101584 may serve as an example of a well-studied post-AGB
binary~[\cite{Bakker1996}]. This system contains a hot B9\,II-type
post-AGB star and a low-mass companion of unknown nature. It may be either
a white dwarf or a low-mass MS star [\cite{Bakker1996}].

\section{Conclusions}\label{conclus}

The results of our quantitative spectral classification of BD$-6\degr$1178
led us to conclude that it is a double-line spectroscopic binary. Both 
components are F-type stars: F5\,IV--III\,+\,F3V. We measured the
heliocentric velocities of both components of the binary at four time
moments. We found no grounds to classify BD$-6\degr$1178 as a post-AGB
star. The coordinates of BD$-6\degr$1178 and its heliocentric
distance 450\,pc allow it to be suspected of a membership in the
Ori\,OB1 association. Thus BD$-6\degr$1178 may be a young pre-MS object of
the Galactic disk.

\section*{Acknowledgements}

We are much indebted to V.E.~Panchuk and M.V.~Yushkin for their
help with observations at the 6-m telescope of the Special
Astrophysical Observatory of the Russian Academy of Sciences. This work
was supported by the Russian Foundation for Basic Research (project
No.\,08-02-00072\,a), the ``Extended Objects in the Universe'' program of
fundamental research of the Division of Physical Sciences of the Russian
Academy of Sciences, and the program ``The Origin and Evolution of Stars
and Galaxies'' of the Presidium of the Russian Academy of Sciences. In this
work data from the SIMBAD and CDS of the Strasbourg Astronomical Data
Center databases were used.

\newpage

\begin{table}
\bigskip
\caption{Log of the observations of  BD$-6\degr$1178, average heliocentric
         radial velocities Vr for both components; the corresponding
	 rotational velocities $V \sin i $ are given in paretheses.}
\begin{tabular}{ r @{\quad} c @{\quad}  c @{\quad}  r   r}
\hline
Date     & JD    & $\Delta \lambda$,\,\AA &\multicolumn{2}{c}{Vr ($V \sin i $), km/s}\\
\cline{4-5}

     &  2453+ &                  & 1-component &  2-component \\
\hline
 9.03.04 &  074.2 &  5300--6770  &  85 (24)  &$-$36 (18)  \\
18.01.05 &  389.2 &  5300--6770  &$-$20 (24) &   66 (19)  \\
24.09.05 &  637.5 &  4190--5520  &  84 (24)  &$-$34 (20)  \\
13.11.05 &  688.4 &  4560--6010  &  21 (25:) &   33: (--) \\
\hline
\end{tabular}
\label{data}
\end{table}

\newpage
\clearpage

\newpage
\clearpage

\newpage
\clearpage

\newpage
\clearpage

\end{document}